\documentclass[referee]{raa}           

\usepackage{graphicx,times}
\usepackage{natbib}
\usepackage{amssymb,amsmath}
\usepackage{soul}
\bibpunct{(}{)}{;}{a}{}{,}

\graphicspath{{./}{figures/}}

\usepackage[pagebackref=true]{hyperref}
\defcitealias{HvVC2023}{H23}

\begin{document}

   \title{Fitting the light curves of tidal disruption events with non-parabolic model}

 \volnopage{ {\bf 20XX} Vol.\ {\bf X} No. {\bf XX}, 000--000}
   \setcounter{page}{1}

   \author{Shiyan Zhong\inst{1}, 
   Chenxi Shang\inst{1,2}, 
   Xiaowei Liu\inst{1}
   }

   \institute{ South-Western Institute for Astronomy Research, Yunnan University, Kunming, 650500 Yunnan, China; {\it zhongsy@ynu.edu.cn}\\
        \and
             Department of Astronomy, Yunnan University, Kunming, 650500 Yunnan, China\\
\vs \no
   {\small Received 20XX Month Day; accepted 20XX Month Day}
}

\abstract{
Tidal disruption events (TDEs) are powerful probes of supermassive black hole (SMBH) properties and accretion physics. The existing light curve fitting tools assume that the disrupted stars are on parabolic orbits, which may introduce systematic biases in derived parameters. In this work, we extend the model of \cite{Zhong2025} to construct a non-parabolic TDE model that incorporates orbital energy of the disrupted star as a free parameter ($\tilde{\epsilon}_{\rm orb}$) to modify the debris mass distribution and mass fallback rate. We apply this model to 30 TDEs from the ZTF-I survey and compare the results with those from a standard parabolic model. We find that neglecting orbital energy leads to biased black hole mass estimates: for eccentric (hyperbolic) orbits, parabolic models systematically underestimate (overestimate) the black hole mass. Additionally, we measure orbital eccentricities ($e$) and penetration factors ($\beta$) of the disrupted stars in this sample, enabling an investigation of their origins via the $e$-$\beta$ parameter space. Most events (24/30) are consistent with production via two-body relaxation in spherical nuclear star clusters, but six outliers with high $\beta$ and $e<1$ suggest alternative mechanisms. Our results highlight the importance of accounting for orbital energy in TDE modeling to improve the accuracy of SMBH mass measurements and to better understand the dynamical origin of the disrupted stars.
}

   \authorrunning{Zhong, Shang, Liu}            
   \titlerunning{Non-parabolic TDE model}  
   \maketitle

%
\section{Introduction}           
\label{SECT:Introduction}
A star comes too close to a supermassive black hole (SMBH) shall be disrupted by the overwhelming tidal force from the SMBH, causing a tidal disruption event (TDE). These events generate flares lasting from months to years \cite{Rees1988}. The critical distance, denoted as the tidal radius $r_{\rm t}$, to the SMBH for destroying a star is typically on the order of a few to tenths of gravitational radius, $r_{\rm g}=GM_{\rm BH}/c^2$, where $G$, $M_{\rm BH}$ and $c$ are the gravitational constant, black hole mass and speed of light, respectively. Because TDEs and the ensuing accretion processes occur so close to the SMBH, they provide a unique probe of black hole properties (particularly mass) and accretion physics.

Currently, two publicly available tools for fitting tidal disruption event (TDE) light curves are widely used: the Modular Open Source Fitter for Transients (\texttt{MOSFiT})~\citep{MOSFiT,MGR2019} and \texttt{TiDE}~\citep{TiDE}. Both aim to extract key physical parameters, such as the supermassive black hole mass ($M_{\rm BH}$), the mass of the disrupted star ($m_*$), and the penetration factor $\beta$ (defined as the ratio of the tidal disruption radius $r_{\rm t}$ to the pericenter distance $r_{\rm p}$), which governs the disruption strength. These tools employ distinct physical models: \texttt{MOSFiT} uses a luminosity-dependent photosphere with a blackbody spectral energy distribution (SED), whereas \texttt{TiDE} incorporates an accretion disk and a reprocessing layer. Both have been applied in numerous studies \citep{MGR2019,Gomez+2020,Nicholl+2020MNRAS,MR2021ApJ,HvVC2023,KV2023,Zhong+2025RAA}.

Both \texttt{MOSFiT} and \texttt{TiDE} assume that the disrupted star is on a parabolic orbit prior to disruption, based on the expectation that such stars originate from the influence radius of the black hole\footnote{By definition, stars within the influence radius are bound to the black hole, while those outside are unbound.}~\citep{FR1976}. We refer to such events as parabolic TDEs in this work.

However, the recent progresses in theoretical and observational studies has demonstrated that the assumption of eccentricity $e=1$ is not always valid. \cite{HZL2018} conducted a series of $N$-body simulations to investigate the $e$ and $\beta$ of the TDEs. They found that almost no TDE can be strictly classified as parabolic. Instead, their simulations revealed that most TDEs fall into the categories of either marginally eccentric or marginally hyperbolic. \cite{HSL2013MNRAS} showed that for $e<1$ orbit the fallback rate can deviate strongly from the standard $t^{-5/3}$ behavior, and that below a critical eccentricity all debris can remain bound and return within a finite time. \cite{PH2020ApJ} discussed the fallback properties over a wider range of orbits, including marginally eccentric, parabolic, and marginally hyperbolic cases, and clarified the dependence on $e$ and $\beta$. Although the deviation of the eccentricity from unity is only of the order of $10^{-3}$, the associated relative changes in the mass fallback rate and the relevant timescales can reach approximately $0.1$ (as detailed in Section~\ref{SUBSEC:BH_mass_vs_Eorb}).

On the other hand, many repeated partial TDEs have been reported in the literature: AT 2018fyk~\citep{Wevers+2023}, AT 2020vdq~\citep{Somalwar+2025}, AT 2022dbl (\citealp{LinJiangWang+2024ApJ,Hinkle+2024,Makrygianni+2025ApJL}), IRAS F01004-2237~\citep{SunJiangDou+2024}, AT 2021aeuk (\citealp{BaoGuoZhang+2024,SunGuoGu+2025}) and AT 2023uqm~\citep{WangWangHuang+2025}. For repeating events to occur, the star must reside on an eccentric orbit; otherwise, it can encounter the central black hole only once. Therefore, the eccentricity (or more fundamentally, the orbital energy) of the disrupted star is a crucial parameter in precise modeling of TDE light curves. However, existing models typically assume the eccentricity to be unity. As we will demonstrate, neglecting eccentricity in light curve modeling can introduce systematic biases in black hole mass measurements (Section~\ref{SUBSEC:BH_mass_vs_Eorb} and \ref{SUBSEC:BH_mass_compare}).

In this work, we employ a modified version of the model presented by \cite{Zhong2025} to fit the light curves of the 30 TDEs observed in the ZTF-I survey~(\citealp{HvVC2023}, hereafter \citetalias{HvVC2023}). This non-parabolic TDE model incorporates the effect of orbital energy of the disrupted star and is described in detail in Section~\ref{SECT:Model}. We present the fitting results in Section~\ref{SECT:RESULTS}, where we compare the black hole masses derived from both parabolic and non-parabolic TDE models.
Our analysis demonstrates that the differences between these mass estimates depend on the orbital energy of the disrupted star. 
Furthermore, our model enables the measurement of the orbital eccentricity; combined with the penetration parameter $\beta$, this allows us to investigate the origin of the disrupted stars using the theoretical framework of \cite{ZHL2023ApJ}. We summarize our findings in Section~\ref{SECT:Summary}.

\section{Non-parabolic TDE Model}
\label{SECT:Model}
\cite{Zhong2025} developed a light curve model to fit repeated partial TDEs produced by stars on eccentric orbits. In this work, we utilize this light curve model with minor modifications to fit 30 TDEs from the ZTF-I survey. We provide a brief description of the model here; further details can be found in Section 2 of \cite{Zhong2025}.

We first retrieve the mass distribution of the tidal debris, $(\mathrm{d} m/\mathrm{d} \epsilon)_{e=1}$ from the library provided by~\citet{GRR2013}, for a given combination of $M_{\rm BH}$, $m_{*}$ and $\beta$, assuming the star is on a parabolic orbit. \cite{GRR2013} adopted two polytropic stellar models: $\gamma=4/3$ and $\gamma=5/3$. For each types of polytropic star, the mass stripping starts at $\beta_{\rm p}$ and full disruption happens at $\beta_{\rm d}$. Note, $\beta_{\rm p}$ and $\beta_{\rm d}$ take different values in different stellar models. To avoid confusion in interpreting the results, we follow \texttt{MOSFiT} to linearly map the conventional $\beta$ to the scaled penetration factor $b$ (see equation 5 of \cite{Zhong2025} for an example): $b=0$ means no disruption, $0<b<1$ means the star is partially disrupted, $b\geq 1$ means the star is completely disrupted, and $b=2$ corresponds to the maximum $\beta$ that is simulated by \cite{GRR2013}. In the work of \cite{Zhong2025}, the prior range for the scaled penetration factor $b$ was restricted to values below $0.95$, as their study focused on partial TDEs. In this work, we relax this constraint to encompass both partial ($0\leq b<1$) and full TDEs ($1\leq b \leq 2$).

The debris mass distribution of the non-parabolic TDE is obtained by shifting $\mathrm{d} m/\mathrm{d} \epsilon$ curve by the specific orbital energy ($\epsilon_{\rm orb}$) of the star prior to disruption~\citep{HSL2013MNRAS},

\begin{equation}
\frac{\mathrm{d} m}{\mathrm{d}\epsilon} (\epsilon)=\left( \frac{\mathrm{d} m}{\mathrm{d}\epsilon}\right)_{e=1} (\epsilon-\epsilon_{\rm orb}).
\end{equation}

\cite{Zhong2025} treated the orbital period of the star before the first disruption ($P_{\rm orb,0}$) as a fitting parameter, as their study focused on repeated partial TDEs exhibiting two distinct flares. Consequently, the orbital period can be roughly estimated by measuring the time interval between the two flares, which helps constrain the range of $P_{\rm orb,0}$. In a potential generated by a point mass, the specific orbital energy ($\epsilon_{\rm orb}$) can be directly calculated from the orbital period using the Kepler's third law.

However, the aforementioned approach is applicable only to eccentric orbits, as parabolic and hyperbolic orbits do not have a defined orbital period. Therefore, in this work, we directly use the specific orbital energy as a fitting parameter. 
This allows our model to accommodate stars on eccentric ($\epsilon_{\rm orb}<0$), parabolic ($\epsilon_{\rm orb}=0$) and hyperbolic ($\epsilon_{\rm orb}>0$) orbits. However, the range of $\epsilon_{\rm orb}$ is not straightforward to determine. Instead, we employ a scaled specific orbital energy ($\tilde{\epsilon}_{\rm orb}$) as the fitting parameter, defined as:

\begin{equation}
\tilde{\epsilon}_{\rm orb} = \frac{\epsilon_{\rm orb}}{\Delta\epsilon},
\label{Eq:Eorb}
\end{equation}
\noindent
where

\begin{equation}
\Delta\epsilon=GM_{\rm BH}r_*/r_{\rm t}^2,
\label{Eq:Delta_E}
\end{equation}
\noindent
is the typical half-width of the debris mass distribution as a function of specific energy. Here, the tidal radius is expressed as $r_{\rm t}=r_*(M_{\rm BH}/m_*)^{1/3}$, with $r_*$ being the radius of the disrupted star. The permitted range of $\tilde{\epsilon}_{\rm orb}$ in the fitting procedure is typically from $-0.5$ to $0.5$, but in some cases the lower limit is extended to $-1$.

The specific orbital energy of the disrupted star can also be expressed as $\epsilon_{\rm orb}=-GM_{\rm BH}/(2a)$, where $a$ is the semi-major axis, given by $a =r_{\rm p}/(1-e)=r_{\rm t}/[\beta(1-e)]$. Substituting this expression for $\epsilon_{\rm orb}$ and equation~\ref{Eq:Delta_E} into equation~\ref{Eq:Eorb}, we obtain:

\begin{equation}
\tilde{\epsilon}_{\rm orb}=-\frac{\beta(1-e)}{2}q^{1/3},
\label{Eq:sEorb}
\end{equation}
\noindent
where $q=M_{\rm BH}/m_*$ is the mass ratio between the BH and the disrupted star.

The debris mass distribution is converted to the mass fallback rate $\dot{m}_{\rm fb}$ via $\dot{m}_{\rm fb}=(\mathrm{d}m/\mathrm{d}\epsilon)(\mathrm{d}\epsilon/\mathrm{d}t)$, where $\mathrm{d}\epsilon/\mathrm{d}t = (1/3)(2\pi G M_{\rm BH})^{2/3}t^{-5/3}$ is derived from Kepler’s third law. 
Then, a viscous delay (controlled by the viscous timescale $t_{\nu}$) is applied to the above fallback rate using equation 7 of \cite{MGR2019}.
This fallback rate is then used as input for the radiation model to generate synthetic multi-band light curves. 

Numerous radiation models have been proposed to explain the origin of UV/optical emission observed in TDEs. The majority of these invoke a reprocessing layer enshrouding a central engine. This central engine could be either a compact (typical size $\sim 2r_{\rm p}$) super-Eddington accretion disk \citep{Dai+2018ApJ,CurdNarayan2019MNRAS,Qiao+2025MNRAS,GuoQiao2026ApJ,ChenQiao2026MNRAS} or various shock regions within an eccentric accretion flow~\citep{Ryu+2023ApJ.957.12R,Huang+2024ApJ,SS2024Natur}. The reprocessing layer absorbs the soft X-ray and extreme-UV photons from the central engine, and re-emit in the UV/optical bands from an extended  photosphere (much larger than the size of the central engine). Alternatively, \cite{LiuCao+2021ApJ} propose that the UV/optical photons originate directly from an eccentric accretion disk.

For simplicity, we adopt the same luminosity-dependent photosphere model as implemented in \texttt{MOSFiT}. The bolometric luminosity is converted from mass fallback rate via $L_{\rm bol}=\eta\dot{m}_{\rm fb}c^2$, where $\eta$ is the radiative efficiency and is treated as a free parameter. For simplicity, $\eta$ is assumed to be constant throughout the TDE evolution; the potential influence of this assumption is discussed in Section~\ref{SUBSECT:Caveat}. The wide range of $\eta$ encompasses various energy conversion processes: spanning from $10^{-3}$ (the minimum value for an eccentric accretion disk, \cite{ZLK2021}), through the sub-percent level characteristic of stream-stream collisions \citep{Jiang_2016ApJ}, up to $0.42$ for the accretion disk around a maximally spinning BH.

The optical SED is generally
described by blackbody radiation with effective temperature $T_{\rm eff}$~\citep{Gezari2021ARAA}. The blackbody photosphere radius is modeled as $R_{\rm ph} = R_{\rm ph0} a_{\rm p}(L_{\rm bol}/L_{\rm Edd})^l$, where $R_{\rm ph0}$ is a normalization of the photosphere radius, $a_{\rm p}$ is the semimajor axis of the material corresponding to the peak fallback rate, $L_{\rm Edd}$ is the Eddington luminosity, and $l$ is the power law index. It follows that the effective temperature evolves as $T_{\rm eff}\propto L_{\rm bol}^{(1-2l)/4}$. In the special case of $l=1/2$, the effective temperature remains constant over time,  consistent with observations in many optical TDEs~\citep{Yao+2023ApJ}. Following \texttt{MOSFiT}, we also incorporate the hydrogen column density ($n_{\rm H}$) to estimate host galaxy extinction in each observed band and use $\sigma$ to account for potential underestimation of observational errors.

We use the python package \texttt{emcee} \citep{emcee2013PASP} to optimize the model fitting by maximizing the log-likelihood function:

\begin{equation}
\ln p = -\frac{1}{2} \sum_{i} \left[ \frac{x_i^2}{\sigma_i^2+\sigma^2} + \ln 2\pi (\sigma_i^2+\sigma^2) \right] + \ln p_{\rm prior}(\Theta),
\end{equation}
\noindent
where $x_i$ represents the residual between the $i$th observation and the model prediction, $\sigma_i$ is the observational error, $\Theta$ denotes the vector of fitting parameters, and $p_{\rm prior}(\Theta)$ is the prior probability distribution. Maximizing this log-likelihood is equivalent to minimizing the $\chi^2$ statistic, defined as $\chi^2= \sum x_i^2/(\sigma_i^2+\sigma^2)$.

The prior distributions for the common parameters are listed in Table~\ref{table-paras}. Additionally, there are event-specific parameters, such as the time of disruption $t_{\rm disrupt}$.

\begin{table}[htbp]
\caption{The fitting parameters used in the light curve modeling.
\label{table-paras}}
\begin{center}
\begin{tabular}{cccc}
  \hline
  Parameter & Prior Distribution Type & Min  &  Max   \\
  \hline
  $M_{\rm BH}~(M_{\odot})$  & Log   & $10^6$  & $10^8$ \\
  $m_{*}~(M_{\odot})$       & Kroupa & $0.1$ & $100$ \\
  $b$ (scaled penetration factor)  & Flat & $0$ & $2$ \\
  $\tilde{\epsilon}_{\rm orb}$ & Flat & $-0.5$ & $0.5$ \\
  $\eta$                  & Log   & $10^{-4}$ & $0.1$ \\
  $R_{\rm ph0}$           & Log    & $10^{-2}$ & $10^2$ \\
  $l$                     & Flat   & $0$ & $4$ \\
  $t_{\nu}$~(days)          & Log    & $10^{-1}$ & $10^2$ \\
  $n_{\rm H} ~(\rm{atom/cm^2})$  & Log    & $10^{15}$ & $10^{22}$ \\
  $\sigma$                &  Flat    & $0$ & $1$ \\
  \hline
\end{tabular}
\end{center}
\textbf{NOTES:}~The first column gives the name of the parameters. The second column indicate the type of prior distribution for each parameters: ``Flat" means the prior is uniformly sampled from the value range; ``Log" means the prior is logrithmically uniformly sampled in the value range; ``Kroupa" means the stellar mass are sampled from the Kroupa initial mass function. The third and fourth columns give the allowed range for each parameter.
\end{table}

Finally, we calculate the eccentricity of the orbit via $e=1+2r_{\rm t}\epsilon_{\rm orb}/(GM_{\rm BH}\beta)$.

\subsection{The influence of orbital energy on the BH mass measurement}
\label{SUBSEC:BH_mass_vs_Eorb}

In this subsection, we first investigate the influence of orbital energy on the peak mass fallback rate and the evolutionary timescale of the fallback process. We then examine how orbital energy affects the black hole and stellar mass measurements.

Consider a simplified debris mass distribution for a parabolic full TDE, where $\mathrm{d}m/\mathrm{d}\epsilon$ is constant with a value of $m_*/(2\Delta\epsilon)$ between $\epsilon_{\rm min}=-\Delta\epsilon$ and $\epsilon_{\rm max}=\Delta\epsilon$. The corresponding fallback rate is given by

\begin{equation}
\frac{\mathrm{d}m}{\mathrm{d}t}
 = \frac{m_*}{3t_{\rm min}}\left( \frac{t}{t_{\rm min}} \right)^{-5/3},~~~(t>t_{\rm min})
\label{Eq:simple_fallback_rate}
\end{equation}
\noindent
where $t_{\rm min}=(\pi/\sqrt{2})GM_{\rm BH}|\epsilon_{\rm min}|^{-1.5}$ is the orbital period of the most bound debris. According to equation~\ref{Eq:simple_fallback_rate}, the peak fallback rate is $\dot{m}_{\rm peak}=m_*/(3t_{\rm min})$.

For non-parabolic TDE, the debris is uniformly distributed between $\epsilon_{\rm min}=\epsilon_{\rm orb}-\Delta\epsilon$ and $\epsilon_{\rm max}=\epsilon_{\rm orb}+\Delta\epsilon$. Consequently, the mass of the bound debris is:

\begin{align*}
m_{\rm bound}
&= \int_{\epsilon_{\rm min}}^{0} \frac{m_* }{2\Delta\epsilon} \mathrm{d}\epsilon \\
  &=\frac{m_*}{2\Delta\epsilon}(\Delta\epsilon-\epsilon_{\rm orb})\\
  &=\frac{m_*}{2}(1-\frac{\epsilon_{\rm orb}}{\Delta\epsilon}).
\label{Eq:simple_fallback_rate}
\end{align*}
\noindent
Inserting equations~\ref{Eq:Eorb} and ~\ref{Eq:sEorb} into the above expression, we get
\begin{equation}
m_{\rm bound} =\frac{m_*}{2}(1-\tilde{\epsilon}_{\rm orb})=\frac{m_*}{2}\left[1+\frac{\beta(1-e)}{2}q^{1/3}\right].
\label{Eq:m_bound}
\end{equation}
\noindent
For a solar-type star disrupted by a $10^6 M_{\odot}$ BH, the cubic root of mass ratio is $q^{1/3}=100$. Adopting typical TDE parameters, $\beta\simeq 1$ and $1-e\simeq \rm{few} \times 10^{-3}$, the fractional difference in bound mass between parabolic and non-parabolic TDEs is order of $0.1$.

From the definition of $t_{\rm min}$, the fractional change in this timescale can be estimated as:

\begin{equation}
    \frac{\Delta t_{\rm min}}{t_{\rm min}}
= 1.5 \frac{\Delta \epsilon_{\rm min}}{\epsilon_{\rm min}}.
\label{Eq:delta_tmin_1}
\end{equation}
\noindent
When comparing parabolic and non-parabolic TDEs, substituting $\Delta \epsilon_{\rm min} = \epsilon_{\rm min,NonPara} - \epsilon_{\rm min,Para} = \epsilon_{\rm orb}$ and $\epsilon_{\rm min}=-\Delta\epsilon$ into equation~\ref{Eq:delta_tmin_1} yields the fractional change in $t_{\rm min}$:
\begin{equation}
    \frac{\Delta t_{\rm min}}{t_{\rm min}}
= 1.5 \left[\frac{\beta(1-e)}{2}q^{1/3}\right],
\label{Eq:delta_tmin_2}
\end{equation}
\noindent
which is again of the order of $0.1$.

For non-parabolic TDEs, inserting $\epsilon_{\rm min}=\epsilon_{\rm orb}-\Delta\epsilon$ into the definition of $t_{\rm min}$ and using equation~\ref{Eq:Eorb} yields:
\begin{equation}
t_{\rm min}=\frac{\sqrt{2}\pi}{2}G^{-1/2}M_{\rm BH}^{1/2}m_*^{-1}r_*^{3/2}(1-\tilde{\epsilon}_{\rm orb})^{-3/2}.
\label{Eq:tmin_sEorb}
\end{equation}
\noindent
And the peak mass fallback rate becomes
\begin{equation}
\dot{m}_{\rm peak}\propto\frac{m_{\rm bound}}{t_{\min}}
=\frac{1}{\sqrt{2}\pi}G^{1/2}M_{\rm BH}^{-1/2}m_*^2 r_*^{-3/2}(1-\tilde{\epsilon}_{\rm orb})^{5/2}.
\label{Eq:mpeak_sEorb}
\end{equation}

The simple fallback rate (equation~\ref{Eq:simple_fallback_rate}) is fully determined by $t_{\rm min}$ and $\dot{m}_{\rm peak}$. We note that both $t_{\rm min}$ and $\dot{m}_{\rm peak}$ vary with the scaled orbital energy $\tilde{\epsilon}_{\rm orb}$ when other parameters are held constant. Thus, under the constraint of identical $t_{\rm min}$ and $\dot{m}_{\rm peak}$, the parabolic and non-parabolic models will yield different estimates for $M_{\rm BH} $ and $m_*$. In the following analysis, we examine how the fitted $M_{\rm BH}$ and $m_*$ change when $t_{\rm min}$ and $\dot{m}_{\rm peak}$ are fixed and $\tilde{\epsilon}_{\rm orb}$ deviates from $0$. Although equation~\ref{Eq:sEorb} suggests that $\tilde{\epsilon}_{\rm orb}$ (as a proxy for $\epsilon_{\rm orb}$) depends on $M_{\rm BH}$ and $m_*$, in practice, $\epsilon_{\rm orb}$ of the disrupted star is determined solely by stellar dynamics in the host nuclear star cluster (NSC). Therefore, $\epsilon_{\rm orb}$ should be independent of $M_{\rm BH}$ and $m_*$. Since we take $\tilde{\epsilon}_{\rm orb}$ as a proxy of $\epsilon_{\rm orb}$ for technical reasons (as previously mentioned), we treat $\tilde{\epsilon}_{\rm orb}$ as an independent quantity and do not further expand it in terms of $M_{\rm BH}$ and $m_*$. This approach also facilitates comparison between our fitting results and the analytical relations derived here (see Section~\ref{SUBSEC:BH_mass_compare}).

Let us assume $M_{\rm BH}\propto (1-\tilde{\epsilon}_{\rm orb})^{x}$, $m_*\propto (1-\tilde{\epsilon}_{\rm orb})^{y}$, and $r_* \propto m_*^{0.8}$. This yields $t_{\rm min} \propto (1-\tilde{\epsilon}_{\rm orb})^{\frac{1}{2}x+\frac{1}{5}y-\frac{3}{2}}$, and $\dot{m}_{\rm peak}\propto (1-\tilde{\epsilon}_{\rm orb})^{-\frac{1}{2}x+\frac{4}{5}y+\frac{5}{2}}$. To ensure that $t_{\rm min}$ and $\dot{m}_{\rm peak}$ remain constant as $\tilde{\epsilon}_{\rm orb}$ varies, we set the exponents to zero:

\begin{align*}
\frac{1}{2}x+\frac{1}{5}y-\frac{3}{2} &= 0, \\
-\frac{1}{2}x+\frac{4}{5}y+\frac{5}{2} &= 0,
\end{align*}
\noindent
Solving this system of equations, we find $x=17/5$, and $y=-1$. Therefore, the values of $M_{\rm BH}$ and $m_*$ must scale as:

\begin{equation}
M_{\rm BH} \propto (1-\tilde{\epsilon}_{\rm orb})^{17/5},
\label{eq:MBH-sEorb-relation}
\end{equation}
\noindent
and
\begin{equation}
m_*\propto (1-\tilde{\epsilon}_{\rm orb})^{-1}.
\label{eq:mstar-sEorb-relation}
\end{equation}
\noindent
In the extreme case of an eccentric TDE with $\tilde{\epsilon}_{\rm orb}=-1$ (where all the debris remain bound to the BH), the best-fit BH mass would be approximately 10 times larger than that fitted under the assumption $\tilde{\epsilon}_{\rm orb}=0$ (note that $1-\tilde{\epsilon}_{\rm orb}$ increases as eccentricity decreases).

In summary, if a TDE is produced by a star on an eccentric (hyperbolic) orbit, fitting its light curve with parabolic TDE model will systematically underestimate (overestimate) the $M_{\rm BH}$ (see Section~\ref{SUBSEC:BH_mass_compare}).


\subsection{Potential influence of constant radiative efficiency assumption}
\label{SUBSECT:Caveat}

In both the current model and \texttt{MOSFiT}, $\eta$ is treated as a constant. However, this assumption may break down in realistic TDEs. Since the total luminosity sums contributions from all energy sources, whose relative contributions evolve over time, $\eta$ should exhibit temporal variation. Focusing solely on the rising phase of the light curve, \cite{SS2024Natur} found that the dissipation rate of the shocks increases with time. Similarly, in the theoretical stream evolution model of \cite{Bonnerot+2017}, the dissipation rate increases with the number of stream self-intersections. Therefore, adopting a constant $\eta$ could bias the fitting results. For instance, if $\eta$ evolves as found by \cite{SS2024Natur}, the mass fallback rate around and after the peak must be reduced to reproduce the observed light curve. This would imply a lower mass for the bound debris, which could be achieved by reducing the stellar mass and/or $\beta$. To quantitatively assess the influence of constant-$\eta$ assumption, one needs to model the temporal evolution of $\eta$, which is beyond the scope of this paper, and we defer this investigation to future study.

\section{Fitting Results and Discussions}
\label{SECT:RESULTS}
\citetalias{HvVC2023} have published the light curves of 30 TDEs observed during the ZTF-I survey. Their dataset includes observations from \textit{Swift/UVOT} and/or ATLAS for some events. They utilized \texttt{MOSFiT} and \texttt{TDEmass} to estimate the masses of the host BHs and the disrupted stars of these TDEs. However, their light curves were binned to $3\sigma$ detections, whereas the machine-readable data contain all photometric points. To reduce computation time, we also applied a binning procedure, though the resulting light curves may differ slightly from theirs. Consequently, the fitting results for the parabolic TDE model are derived using our own code by fixing $\tilde{\epsilon}_{\rm orb}=0$.

The fitting results from the non-parabolic and parabolic TDE models are presented in Table~\ref{table-results-NP_model} and Table~\ref{table-results-P_model}, respectively. We list only the parameters of primary interest: $M_{\rm BH}$, $m_*$ and the two orbital parameters (provided in both scaled and standard forms).

To evaluate which model provides a better fit to the light curves, we also report the Watanabe-Akaike Information Criterion (WAIC) score in Table~\ref{table-results-NP_model} and Table~\ref{table-results-P_model}. The WAIC score is computed following the same methodology as in \texttt{MOSFiT}, such that a higher WAIC value indicates a better fit. Our analysis reveals that for some events, the non-parabolic model yields a higher WAIC score, while for others, the parabolic model performs better. However, the differences in WAIC scores are generally not large enough to rule out the alternative model in most cases. Additionally, visual inspections of the fitted light curves show that the non-parabolic TDE model performs substantially better for AT 2019meg and AT 2020riz. This suggests that while accounting for orbital eccentricity can improve the fit for certain TDEs, the improvement is often marginal with current data and models. Consequently, both models remain viable for a substantial fraction of the sample. We suggest that future improvements in both light curve observations and TDE theoretical models are needed to better characterize these events.

\subsection{BH mass comparison}
\label{SUBSEC:BH_mass_compare}

\begin{figure}
    \centering
    \includegraphics[width=0.8\linewidth]{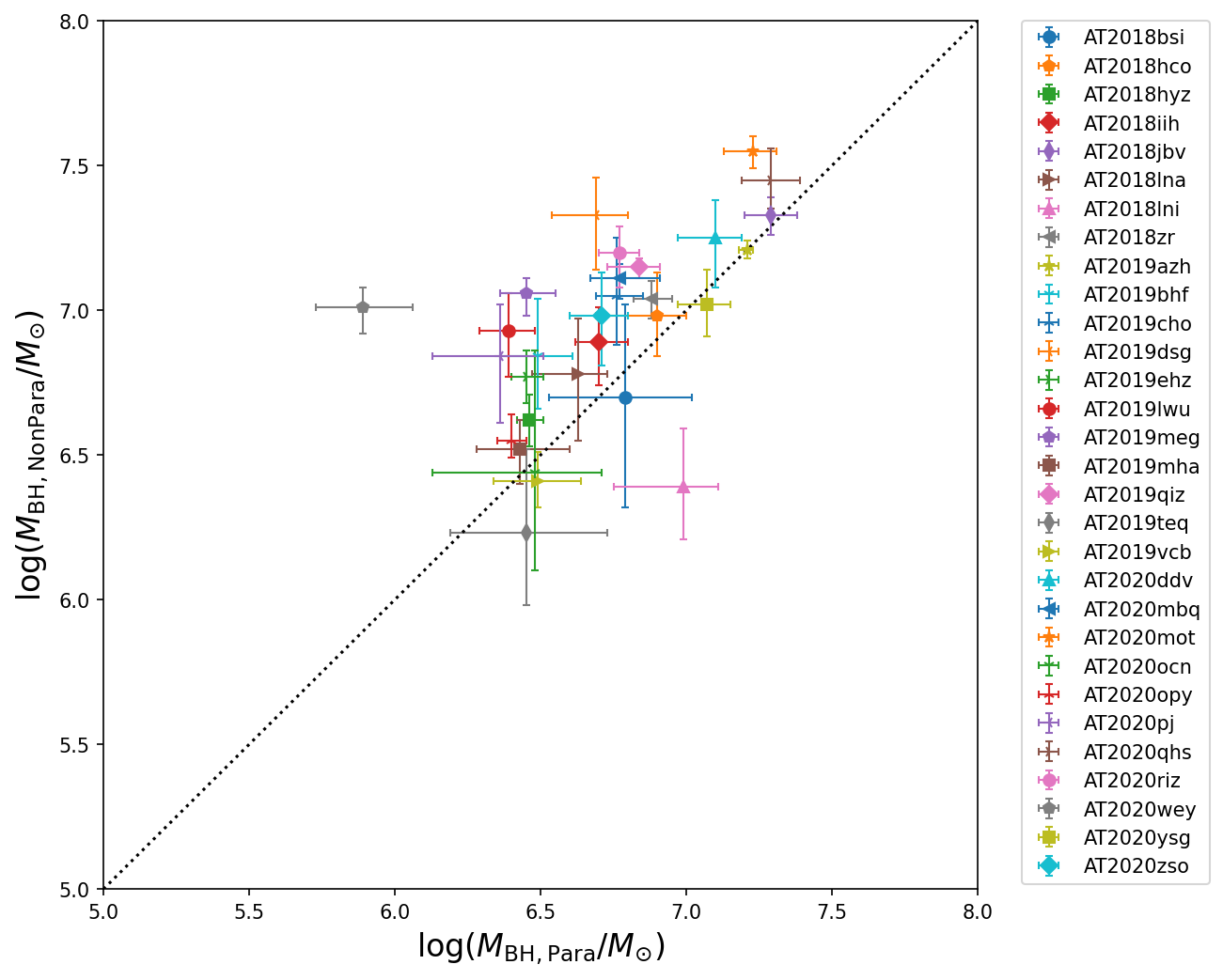}
    \caption{Comparison of BH masses obtained from parabolic TDE model ($M_{\rm BH,Para}$) and non-parabolic TDE model ($M_{\rm BH,NonPara}$). The dotted line is the 1:1 line. The error bar indicates $1\sigma$ uncertainty.
    }
    \label{fig:MBH_comparison}
\end{figure}

We compare the BH masses of individual TDEs fitted by non-parabolic model ($M_{\rm BH,NonPara}$) and parabolic model ($M_{\rm BH,Para}$) in Figure~\ref{fig:MBH_comparison}. We observe that for most of the TDEs, the BH masses obtained with non-parabolic TDE model show significant deviation from the 1:1 line. This is not surprising since in Section~\ref{SUBSEC:BH_mass_vs_Eorb} we learn that the non-parabolic TDE model should report different BH mass from the parabolic TDE model.

\begin{figure}
    \centering
    \includegraphics[width=0.8\linewidth]{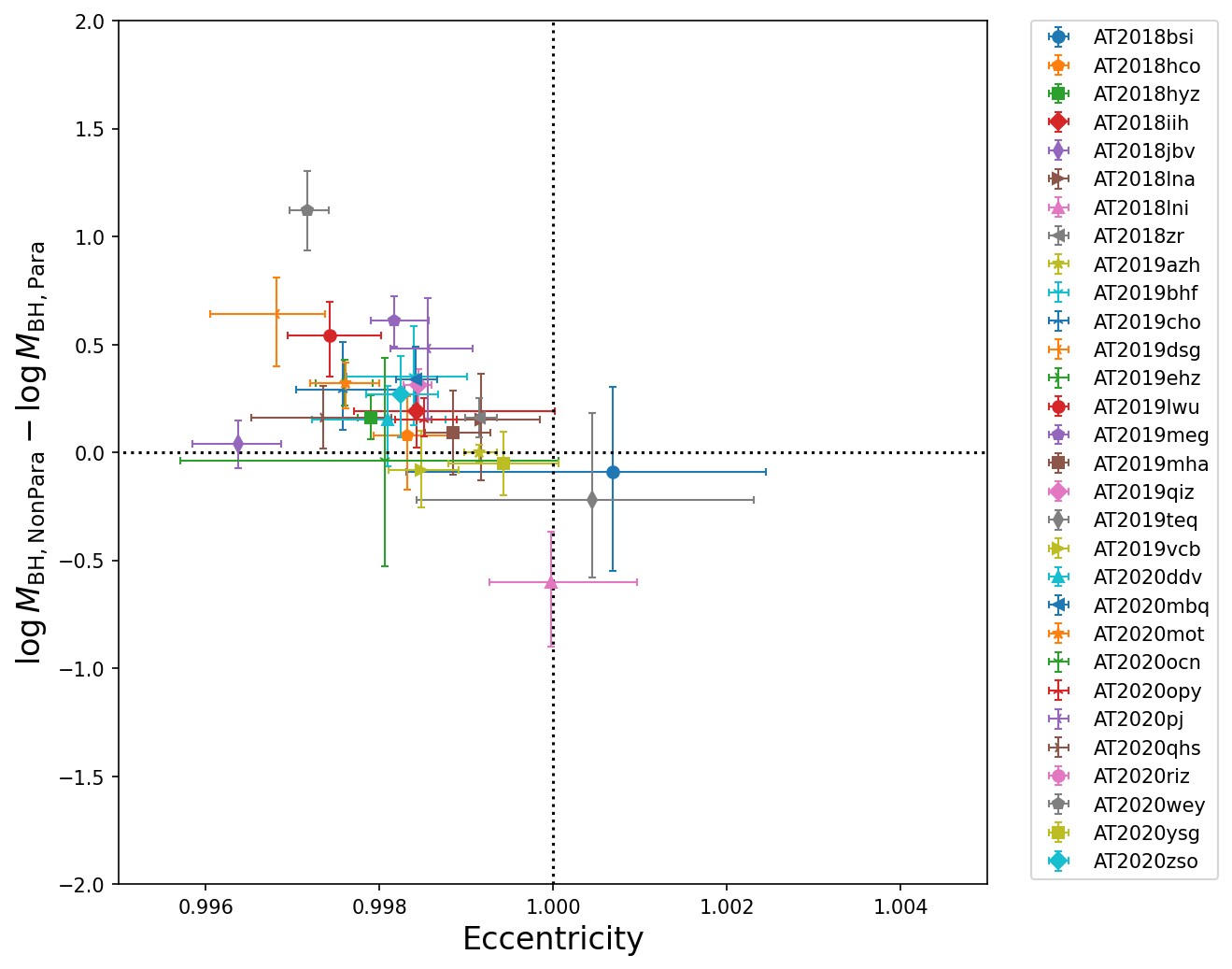}
    \includegraphics[width=0.8\linewidth]{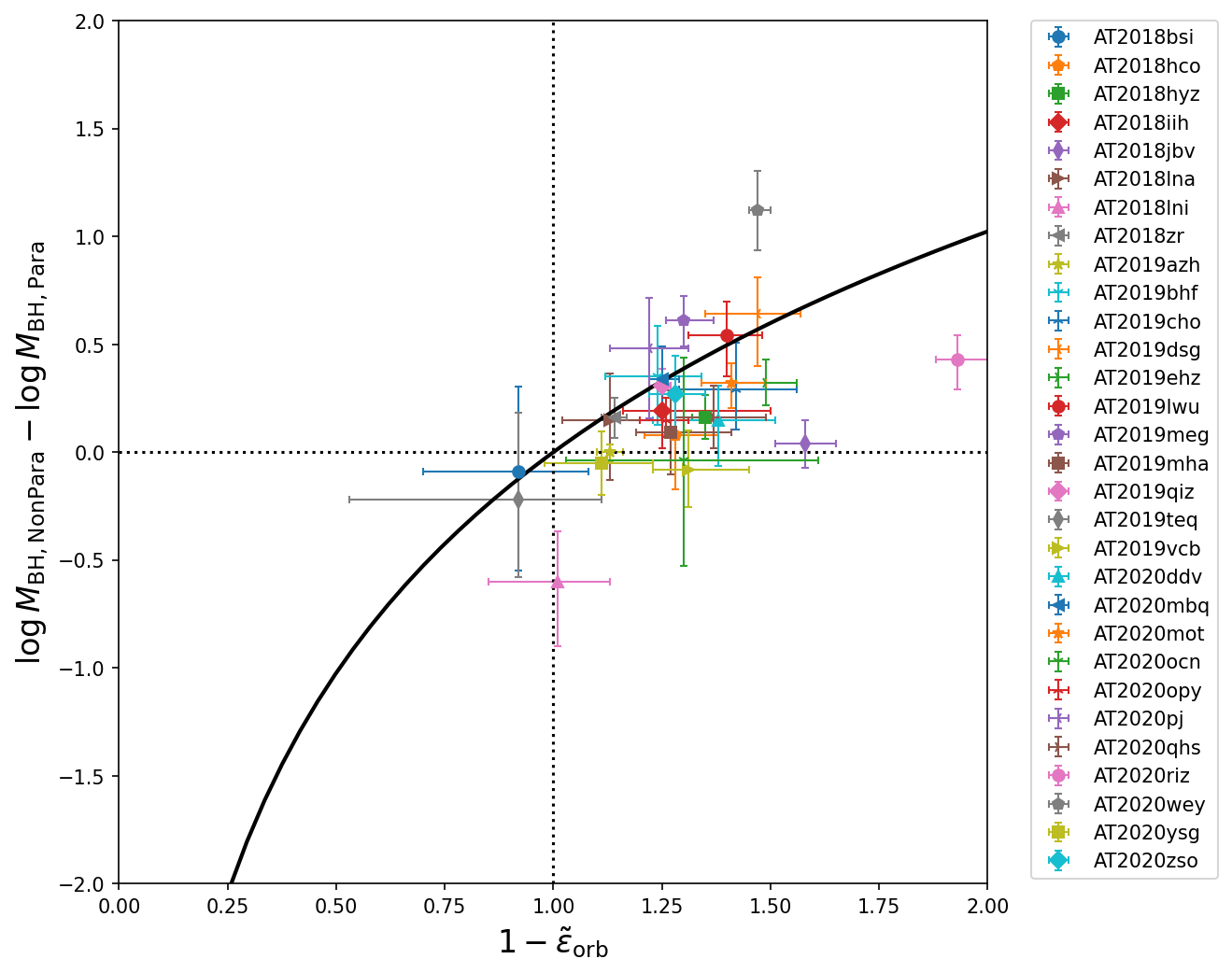}
    \caption{The dependence of the fitted BH mass difference on the orbital eccentricity (upper panel) and $1-\tilde{\epsilon}_{\rm orb}$ (lower panel). The black curve in the lower panel represents the $M_{\rm BH} \propto (1-\tilde{\epsilon}_{\rm orb})^{17/5}$ relation derived in Section~\ref{SUBSEC:BH_mass_vs_Eorb}.
    }
    \label{fig:DeltaMBH_ecc}
\end{figure}

To further investigate the reason for the deviation in BH mass measurements, we plot the quantity $\Delta \log M_{\rm BH}$ ($= \log M_{\rm BH,NonPara}-\log M_{\rm BH,Para}$) against the orbital eccentricity $e$ and $1-\tilde{\epsilon}_{\rm orb}$ in Figure~\ref{fig:DeltaMBH_ecc}. The upper panel of Figure~\ref{fig:DeltaMBH_ecc} clearly shows that $\Delta \log M_{\rm BH}$ decreases with increasing eccentricity, and for most of the $e<1$ TDEs, $\Delta \log M_{\rm BH}>0$, which means for these TDEs parabolic model underestimates the BH mass. On the other hand, the two TDEs with $e>1$ have $\Delta \log M_{\rm BH}<0$ as expected, although the error bars are large. According to the sample TDEs presented in this work, $90\%$ of TDEs have $e<1$ (the reason for the scarcity of $e>1$ TDEs is explained in Section~\ref{SUBSEC:e-beta-distribution}), we suggest that most of the MBH masses measured with parabolic model might underestimate their values.

The lower panel of Figure~\ref{fig:DeltaMBH_ecc} shows the dependence of $\Delta \log M_{\rm BH}$ on the scaled orbital energy $\tilde{\epsilon}_{\rm orb}$. We also plot the $M_{\rm BH} \propto (1-\tilde{\epsilon}_{\rm orb})^{17/5}$ relation derived in Section~\ref{SUBSEC:BH_mass_vs_Eorb} for reference. The 30 TDEs generally follows the relation, although with large scatter. These deviations could come from two sources. First, the equation~\ref{eq:MBH-sEorb-relation} is derived for the simple fallback rate, while in the actual fitting procedure, the mass fallback rate is more complicated, for example, its shape also depend on $\beta$ \citep{GRR2013}, which is not considered. Second, the parameters related to the radiation process are not considered in deriving equation~\ref{eq:MBH-sEorb-relation}. For example, given the same bolometric light curve, the variation of $\eta$ could result in different $\dot{m}_{\rm peak}$ of the mass fallback curve.

\begin{table}[htbp]
\caption{The fitting results from non-parabolic TDE model.}
\label{table-results-NP_model}
\begin{center}
\begin{tabular}{c|c|c|cc|cc|c}
  \hline
  Name & $\log(M_{\rm BH}/M_{\odot})$ & $m_{*}/M_{\odot}$  &  $\tilde{\epsilon}_{\rm orb}$ & $e$ & $b$  & $\beta$ & WAIC \\
  \hline
AT2018bsi & $6.70^{+0.32}_{-0.38}$ & $0.22^{+0.12}_{-0.09}$ & $0.08^{+0.16}_{-0.22}$ & $1.0007^{+0.0018}_{-0.0024}$ & $0.54^{+0.41}_{-0.31}$ & $0.73^{+0.16}_{-0.13}$ & $-34.7$ \\
AT2018hco & $6.98^{+0.15}_{-0.14}$ & $0.24^{+0.10}_{-0.07}$ & $-0.28^{+0.10}_{-0.07}$ & $0.9983^{+0.0006}_{-0.0004}$ & $0.90^{+0.24}_{-0.14}$ & $0.97^{+0.15}_{-0.16}$ & $51.8$ \\
AT2018hyz & $6.62^{+0.09}_{-0.09}$ & $0.35^{+0.05}_{-0.11}$ & $-0.35^{+0.03}_{-0.03}$ & $0.9979^{+0.0002}_{-0.0001}$ & $1.30^{+0.06}_{-0.05}$ & $1.47^{+0.09}_{-0.10}$ & $98.8$ \\
AT2018iih & $6.89^{+0.12}_{-0.15}$ & $0.22^{+0.08}_{-0.07}$ & $-0.25^{+0.25}_{-0.09}$ & $0.9984^{+0.0016}_{-0.0007}$ & $1.00^{+0.05}_{-0.03}$ & $0.91^{+0.07}_{-0.02}$ & $74.8$ \\
AT2018jbv & $7.33^{+0.06}_{-0.07}$ & $11.37^{+2.97}_{-3.58}$ & $-0.58^{+0.07}_{-0.07}$ & $0.9964^{+0.0005}_{-0.0005}$ & $1.31^{+0.05}_{-0.03}$ & $2.50^{+0.10}_{-0.06}$ & $52.6$ \\
AT2018lna & $6.78^{+0.19}_{-0.23}$ & $0.21^{+0.12}_{-0.07}$ & $-0.13^{+0.11}_{-0.11}$ & $0.9992^{+0.0007}_{-0.0006}$ & $1.00^{+0.26}_{-0.29}$ & $0.94^{+0.43}_{-0.15}$ & $17.4$ \\
AT2018lni & $6.39^{+0.20}_{-0.18}$ & $0.24^{+0.15}_{-0.09}$ & $-0.01^{+0.12}_{-0.16}$ & $1.0000^{+0.0010}_{-0.0007}$ & $1.40^{+0.36}_{-0.31}$ & $1.58^{+0.60}_{-0.49}$ & $44.6$ \\
AT2018zr & $7.04^{+0.06}_{-0.07}$ & $0.27^{+0.03}_{-0.05}$ & $-0.14^{+0.03}_{-0.03}$ & $0.9991^{+0.0002}_{-0.0002}$ & $1.02^{+0.04}_{-0.07}$ & $0.93^{+0.06}_{-0.05}$ & $22.6$ \\
AT2019azh & $7.21^{+0.03}_{-0.03}$ & $0.30^{+0.03}_{-0.02}$ & $-0.13^{+0.03}_{-0.03}$ & $0.9991^{+0.0002}_{-0.0002}$ & $0.76^{+0.11}_{-0.08}$ & $0.82^{+0.04}_{-0.04}$ & $-46.4$ \\
AT2019bhf & $6.84^{+0.20}_{-0.18}$ & $0.22^{+0.10}_{-0.08}$ & $-0.24^{+0.10}_{-0.12}$ & $0.9984^{+0.0006}_{-0.0008}$ & $0.93^{+0.18}_{-0.20}$ & $0.88^{+0.22}_{-0.08}$ & $-26.4$ \\
AT2019cho & $7.05^{+0.20}_{-0.17}$ & $1.31^{+0.94}_{-0.49}$ & $-0.42^{+0.14}_{-0.13}$ & $0.9976^{+0.0007}_{-0.0005}$ & $0.93^{+0.11}_{-0.10}$ & $1.70^{+0.20}_{-0.17}$ & $39.5$ \\
AT2019dsg & $7.33^{+0.13}_{-0.19}$ & $1.07^{+0.32}_{-0.10}$ & $-0.47^{+0.10}_{-0.12}$ & $0.9968^{+0.0006}_{-0.0008}$ & $0.41^{+0.06}_{-0.05}$ & $1.10^{+0.08}_{-0.07}$ & $-19.5$ \\
AT2019ehz & $6.77^{+0.09}_{-0.09}$ & $0.43^{+0.05}_{-0.08}$ & $-0.49^{+0.07}_{-0.07}$ & $0.9976^{+0.0003}_{-0.0003}$ & $1.37^{+0.10}_{-0.11}$ & $1.68^{+0.19}_{-0.19}$ & $65.8$ \\
AT2019lwu & $6.93^{+0.13}_{-0.16}$ & $0.24^{+0.09}_{-0.09}$ & $-0.40^{+0.08}_{-0.09}$ & $0.9974^{+0.0006}_{-0.0005}$ & $1.00^{+0.04}_{-0.06}$ & $0.92^{+0.10}_{-0.04}$ & $-26.7$ \\
AT2019meg & $7.06^{+0.05}_{-0.08}$ & $0.34^{+0.07}_{-0.08}$ & $-0.30^{+0.07}_{-0.04}$ & $0.9982^{+0.0004}_{-0.0003}$ & $1.02^{+0.01}_{-0.02}$ & $0.98^{+0.12}_{-0.08}$ & $3.0$ \\
AT2019mha & $6.52^{+0.10}_{-0.12}$ & $0.28^{+0.10}_{-0.09}$ & $-0.27^{+0.14}_{-0.08}$ & $0.9989^{+0.0004}_{-0.0003}$ & $1.62^{+0.10}_{-0.36}$ & $1.91^{+0.26}_{-0.56}$ & $23.0$ \\
AT2019qiz & $7.15^{+0.03}_{-0.04}$ & $0.30^{+0.02}_{-0.03}$ & $-0.25^{+0.02}_{-0.03}$ & $0.9984^{+0.0001}_{-0.0002}$ & $0.90^{+0.10}_{-0.07}$ & $0.87^{+0.04}_{-0.03}$ & $-33.7$ \\
AT2019teq & $6.23^{+0.29}_{-0.25}$ & $0.23^{+0.14}_{-0.09}$ & $0.08^{+0.19}_{-0.39}$ & $1.0005^{+0.0019}_{-0.0020}$ & $1.34^{+0.37}_{-0.23}$ & $1.49^{+0.58}_{-0.40}$ & $15.1$ \\
AT2019vcb & $6.41^{+0.10}_{-0.09}$ & $0.24^{+0.13}_{-0.09}$ & $-0.31^{+0.14}_{-0.08}$ & $0.9985^{+0.0004}_{-0.0004}$ & $1.45^{+0.30}_{-0.25}$ & $1.66^{+0.47}_{-0.39}$ & $28.2$ \\
AT2020ddv & $7.25^{+0.13}_{-0.17}$ & $1.51^{+0.72}_{-0.25}$ & $-0.38^{+0.13}_{-0.15}$ & $0.9981^{+0.0007}_{-0.0009}$ & $0.93^{+0.21}_{-0.20}$ & $1.76^{+0.39}_{-0.25}$ & $-18.9$ \\
AT2020mbq & $7.11^{+0.05}_{-0.07}$ & $0.29^{+0.05}_{-0.06}$ & $-0.25^{+0.04}_{-0.03}$ & $0.9984^{+0.0003}_{-0.0002}$ & $0.92^{+0.11}_{-0.12}$ & $0.88^{+0.10}_{-0.05}$ & $36.4$ \\
AT2020mot & $7.55^{+0.05}_{-0.06}$ & $1.54^{+0.51}_{-0.19}$ & $-0.41^{+0.08}_{-0.07}$ & $0.9976^{+0.0004}_{-0.0004}$ & $0.51^{+0.05}_{-0.04}$ & $1.24^{+0.06}_{-0.05}$ & $53.6$ \\
AT2020ocn & $6.44^{+0.42}_{-0.34}$ & $0.23^{+0.14}_{-0.10}$ & $-0.30^{+0.31}_{-0.27}$ & $0.9981^{+0.0020}_{-0.0024}$ & $1.03^{+0.62}_{-0.57}$ & $1.04^{+0.94}_{-0.35}$ & $-21.2$ \\
AT2020opy & $6.55^{+0.09}_{-0.06}$ & $0.20^{+0.16}_{-0.07}$ & $-0.26^{+0.05}_{-0.06}$ & $0.9985^{+0.0004}_{-0.0003}$ & $1.27^{+0.09}_{-0.07}$ & $1.34^{+0.20}_{-0.11}$ & $83.4$ \\
AT2020pj & $6.84^{+0.18}_{-0.23}$ & $0.24^{+0.14}_{-0.09}$ & $-0.22^{+0.09}_{-0.09}$ & $0.9986^{+0.0005}_{-0.0004}$ & $1.00^{+0.23}_{-0.27}$ & $0.97^{+0.35}_{-0.17}$ & $5.0$ \\
AT2020qhs & $7.45^{+0.11}_{-0.10}$ & $6.52^{+5.20}_{-3.46}$ & $-0.37^{+0.12}_{-0.09}$ & $0.9973^{+0.0008}_{-0.0008}$ & $0.86^{+0.05}_{-0.09}$ & $1.67^{+0.06}_{-0.11}$ & $6.1$ \\
AT2020riz & $7.20^{+0.09}_{-0.12}$ & $3.40^{+2.57}_{-1.23}$ & $-0.93^{+0.09}_{-0.05}$ & $0.9895^{+0.0016}_{-0.0019}$ & $0.38^{+0.09}_{-0.10}$ & $1.08^{+0.11}_{-0.13}$ & $-9.0$ \\
AT2020wey & $7.01^{+0.07}_{-0.09}$ & $0.24^{+0.05}_{-0.06}$ & $-0.47^{+0.03}_{-0.02}$ & $0.9972^{+0.0002}_{-0.0002}$ & $1.03^{+0.02}_{-0.04}$ & $0.95^{+0.04}_{-0.05}$ & $-45.2$ \\
AT2020ysg & $7.02^{+0.12}_{-0.11}$ & $1.17^{+0.59}_{-0.20}$ & $-0.11^{+0.12}_{-0.13}$ & $0.9994^{+0.0006}_{-0.0006}$ & $0.95^{+0.31}_{-0.25}$ & $1.74^{+0.63}_{-0.30}$ & $18.4$ \\
AT2020zso & $6.98^{+0.15}_{-0.17}$ & $0.27^{+0.07}_{-0.11}$ & $-0.28^{+0.07}_{-0.06}$ & $0.9982^{+0.0004}_{-0.0004}$ & $0.94^{+0.22}_{-0.30}$ & $0.89^{+0.27}_{-0.13}$ & $-24.1$ \\
  \hline
\end{tabular}
\end{center}
\textbf{NOTES:} First column is the name of TDE, second column is the MBH mass, third column is the mass of the disrupted star, fourth column is the scaled orbital energy of the disrupted star, fifth column is the orbital eccentricity, sixth column is the scaled penetration factor $b$ and seventh column is the penetration factor. The eighth column is the WAIC score. We use the 16th and 84th percentiles of the posterior distribution to indicate the $1\sigma$ uncertainty in all quantities.
\end{table}

\begin{table}[htbp]
\caption{The fitting results from parabolic TDE model.}
\label{table-results-P_model}
\begin{center}
\begin{tabular}{c|c|c|cc|c}
  \hline
  Name & $\log(M_{\rm BH}/M_{\odot})$ & $m_{*}/M_{\odot}$  &  $b$  & $\beta$ & WAIC\\
  \hline
AT2018bsi & $6.79^{+0.23}_{-0.26}$ & $0.24^{+0.13}_{-0.10}$ & $0.38^{+0.57}_{-0.20}$ & $0.73^{+0.16}_{-0.13}$ & $-32.8$ \\
AT2018hco & $6.90^{+0.10}_{-0.21}$ & $0.44^{+0.05}_{-0.27}$ & $0.86^{+0.09}_{-0.11}$ & $0.97^{+0.15}_{-0.16}$ & $48.8$ \\
AT2018hyz & $6.46^{+0.05}_{-0.04}$ & $0.10^{+0.00}_{-0.00}$ & $0.68^{+0.05}_{-0.05}$ & $1.47^{+0.09}_{-0.10}$ & $85.0$ \\
AT2018iih & $6.70^{+0.10}_{-0.08}$ & $1.30^{+0.74}_{-0.61}$ & $1.03^{+0.04}_{-0.03}$ & $0.91^{+0.07}_{-0.02}$ & $60.5$ \\
AT2018jbv & $7.29^{+0.09}_{-0.09}$ & $8.44^{+4.37}_{-3.95}$ & $0.64^{+0.07}_{-0.08}$ & $2.50^{+0.10}_{-0.06}$ & $50.9$ \\
AT2018lna & $6.63^{+0.10}_{-0.16}$ & $0.20^{+0.13}_{-0.07}$ & $0.86^{+0.22}_{-0.20}$ & $0.94^{+0.43}_{-0.15}$ & $17.3$ \\
AT2018lni & $6.99^{+0.12}_{-0.24}$ & $0.24^{+0.09}_{-0.07}$ & $0.82^{+0.26}_{-0.16}$ & $1.58^{+0.60}_{-0.49}$ & $48.5$ \\
AT2018zr & $6.88^{+0.07}_{-0.06}$ & $0.22^{+0.14}_{-0.09}$ & $0.94^{+0.06}_{-0.09}$ & $0.93^{+0.06}_{-0.05}$ & $17.6$ \\
AT2019azh & $7.21^{+0.02}_{-0.03}$ & $0.44^{+0.03}_{-0.04}$ & $0.76^{+0.08}_{-0.05}$ & $0.82^{+0.04}_{-0.04}$ & $-50.1$ \\
AT2019bhf & $6.49^{+0.12}_{-0.13}$ & $0.17^{+0.13}_{-0.06}$ & $0.82^{+0.16}_{-0.18}$ & $0.88^{+0.22}_{-0.08}$ & $-28.4$ \\
AT2019cho & $6.76^{+0.09}_{-0.07}$ & $0.14^{+0.12}_{-0.03}$ & $0.88^{+0.12}_{-0.16}$ & $1.70^{+0.20}_{-0.17}$ & $45.5$ \\
AT2019dsg & $6.69^{+0.11}_{-0.15}$ & $0.99^{+0.06}_{-0.12}$ & $0.51^{+0.07}_{-0.13}$ & $1.10^{+0.08}_{-0.07}$ & $-32.3$ \\
AT2019ehz & $6.45^{+0.06}_{-0.05}$ & $1.04^{+0.05}_{-0.03}$ & $0.75^{+0.19}_{-0.04}$ & $1.68^{+0.19}_{-0.19}$ & $56.9$ \\
AT2019lwu & $6.39^{+0.09}_{-0.10}$ & $0.11^{+0.07}_{-0.01}$ & $0.76^{+0.18}_{-0.23}$ & $0.92^{+0.10}_{-0.04}$ & $-36.4$ \\
AT2019meg & $6.45^{+0.10}_{-0.09}$ & $0.36^{+0.65}_{-0.26}$ & $1.00^{+0.02}_{-0.03}$ & $0.98^{+0.12}_{-0.08}$ & $-60.3$ \\
AT2019mha & $6.43^{+0.17}_{-0.15}$ & $1.08^{+0.20}_{-0.11}$ & $0.74^{+0.14}_{-0.12}$ & $1.91^{+0.26}_{-0.56}$ & $14.8$ \\
AT2019qiz & $6.84^{+0.07}_{-0.11}$ & $0.40^{+0.04}_{-0.05}$ & $0.98^{+0.04}_{-0.10}$ & $0.87^{+0.04}_{-0.03}$ & $-58.9$ \\
AT2019teq & $6.45^{+0.28}_{-0.26}$ & $0.21^{+0.14}_{-0.08}$ & $1.20^{+0.50}_{-0.44}$ & $1.49^{+0.58}_{-0.40}$ & $15.8$ \\
AT2019vcb & $6.49^{+0.15}_{-0.15}$ & $0.34^{+0.12}_{-0.17}$ & $0.84^{+0.15}_{-0.18}$ & $1.66^{+0.47}_{-0.39}$ & $34.1$ \\
AT2020ddv & $7.10^{+0.09}_{-0.13}$ & $0.35^{+0.10}_{-0.16}$ & $0.75^{+0.17}_{-0.13}$ & $1.76^{+0.39}_{-0.25}$ & $-10.2$ \\
AT2020mbq & $6.77^{+0.14}_{-0.10}$ & $0.20^{+0.27}_{-0.08}$ & $0.86^{+0.11}_{-0.12}$ & $0.88^{+0.10}_{-0.05}$ & $25.5$ \\
AT2020mot & $7.23^{+0.08}_{-0.10}$ & $5.65^{+5.52}_{-2.55}$ & $1.30^{+0.01}_{-0.01}$ & $1.24^{+0.06}_{-0.05}$ & $31.2$ \\
AT2020ocn & $6.48^{+0.23}_{-0.35}$ & $0.21^{+0.20}_{-0.10}$ & $0.47^{+0.87}_{-0.20}$ & $1.04^{+0.94}_{-0.35}$ & $-20.9$ \\
AT2020opy & $6.40^{+0.05}_{-0.05}$ & $1.02^{+0.03}_{-0.02}$ & $0.91^{+0.07}_{-0.04}$ & $1.34^{+0.20}_{-0.11}$ & $63.7$ \\
AT2020pj & $6.36^{+0.15}_{-0.23}$ & $0.16^{+0.11}_{-0.04}$ & $0.93^{+0.23}_{-0.22}$ & $0.97^{+0.35}_{-0.17}$ & $-4.8$ \\
AT2020qhs & $7.29^{+0.10}_{-0.10}$ & $7.67^{+5.07}_{-3.40}$ & $0.62^{+0.11}_{-0.13}$ & $1.67^{+0.06}_{-0.11}$ & $1.1$ \\
AT2020riz & $6.77^{+0.07}_{-0.07}$ & $1.01^{+0.05}_{-0.03}$ & $0.69^{+0.11}_{-0.12}$ & $1.08^{+0.11}_{-0.13}$ & $-47.1$ \\
AT2020wey & $5.89^{+0.17}_{-0.16}$ & $0.12^{+0.11}_{-0.02}$ & $1.04^{+0.11}_{-0.13}$ & $0.95^{+0.04}_{-0.05}$ & $-63.9$ \\
AT2020ysg & $7.07^{+0.08}_{-0.10}$ & $0.28^{+0.07}_{-0.07}$ & $0.89^{+0.17}_{-0.15}$ & $1.74^{+0.63}_{-0.30}$ & $25.3$ \\
AT2020zso & $6.71^{+0.09}_{-0.11}$ & $0.44^{+0.04}_{-0.07}$ & $0.87^{+0.12}_{-0.15}$ & $1.38^{+0.14}_{-0.17}$ & $-31.2$ \\
  \hline
\end{tabular}
\end{center}
\textbf{NOTES:} The columns are the same as Table~\ref{table-results-NP_model}, except for the scaled orbital energy and orbital eccentricity (which are not listed here): they are fixed at zero and unity, respectively, in the parabolic TDE model.
\end{table}

\subsection{The $e$-$\beta$ distribution of the sample TDEs}
\label{SUBSEC:e-beta-distribution}
\begin{figure}
    \centering
    \includegraphics[width=0.8\linewidth]{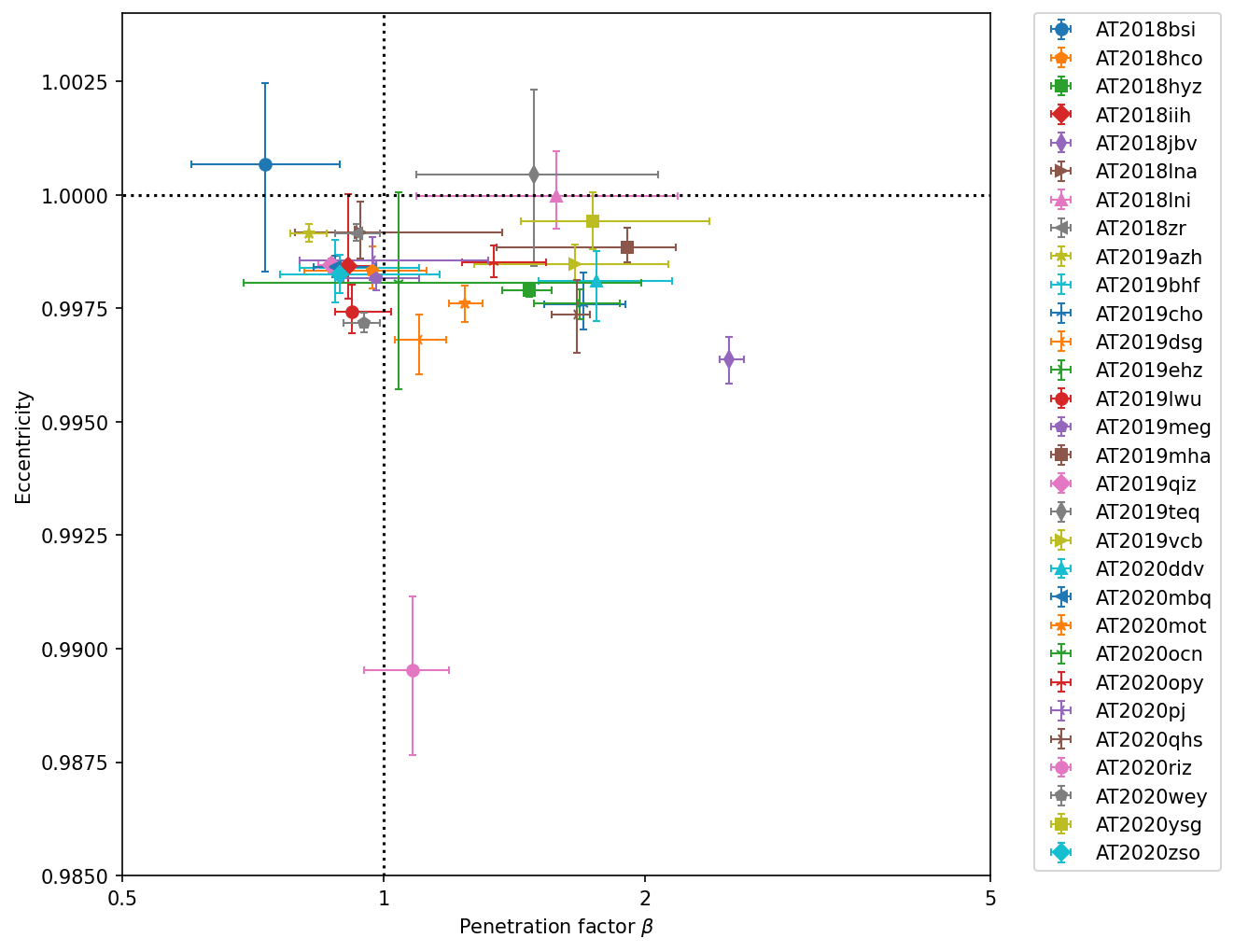}
    \caption{The locations of the 30 TDEs on the $e$-$\beta$ plane.
    }
    \label{fig:ecc-beta}
\end{figure}

Figure~\ref{fig:ecc-beta} shows the distribution of TDEs in the $e$-$\beta$ parameter space. Most events fall within the $e<1$ quadrant, indicating eccentric orbits. Two cases (AT 2018bsi and AT 2019teq) have median eccentricities slightly exceeding unity, though with large uncertainties. The deviation from parabolicity ($1-e$) for most TDEs is of the order of $10^{-3}$, consistent with the findings of \cite{HZL2018} and \cite{ZHL2023ApJ}. The most extreme case (AT 2020riz) exhibits $1-e\simeq 10^{-2}$.

In the following, we explain that the scarcity of $e>1$ TDEs is not due to biases in the non-parabolic TDE model. \cite{ZHL2023ApJ} found that all of the disrupted stars should have orbital energy $\epsilon_{\rm orb} < E_{\rm max}$, where $E_{\rm max}$ is derived as follows. Consider a star falling from infinity toward the central black hole: its specific orbital energy (including the host cluster potential,  $\phi_{\rm cluster}<0$ ) is given by
$v^2/2 - GM_{\rm BH}/r + \phi_{\rm cluster}(r)=0$.
This star has the maximum orbital energy among cluster stars. In this work, $\epsilon_{\rm orb}$ is defined as $v^2/2 - GM_{\rm BH}/r$, so for this star $\epsilon_{\rm orb} = -\phi_{\rm cluster}(r)$. When this star enters the tidal disruption radius, $\epsilon_{\rm orb} = -\phi_{\rm cluster}(r_t) \simeq  -\phi_{\rm cluster}(0)$, and thus $E_{\rm max}\simeq -\phi_{\rm cluster}(0)$. From equation 23 of \cite{ZHL2023ApJ}, the maximum eccentricity is
\begin{equation}
e_{\rm max} = 1+\frac{2}{\beta}\frac{E_{\rm max}}{|E_{\rm t}|}.
\end{equation}
\noindent
where $E_{\rm t}=-GM_{\rm BH}/(2r_{\rm t})$ is the orbital energy for circular orbit at radius $r_{\rm t}$. In most cases, $E_{\rm max}/|E_{\rm t}|$ is of the order of $10^{-4}$, implying that the eccentricity of TDEs cannot exceed $1+$ few $\times 10^{-4}$.

\cite{ZHL2023ApJ} also established a boundary for full TDEs in the $e<1$ quadrant of the $e$-$\beta$ parameter space, based on the work of \cite{CK1978}. TDEs with eccentricities further below unity (i.e. more bound to the MBH) are more likely to originate from the empty loss cone regime. In this regime, the orbit-averaged change in squared angular momentum ($\langle\Delta \mathit{J}^2\rangle$) is small compared to the squared loss cone angular momentum ($\mathit{J}_{\rm lc}^2$), and stars approach the loss cone boundary via diffusion (\citealt{FR1976,LS1977}). From equations (A2) and (A3) of \cite{ZHL2023ApJ}, we have $\beta=r_{\rm t}/r_{\rm p}\simeq J_{\rm lc}^2/J^2$. While many studies (including this work) adopt $r_{\rm t}=r_*(M_{\rm BH}/m_*)^{1/3}$ as the tidal radius for full disruption, hydrodynamic simulations indicate that the true radius for full tidal disruption should be corrected to $r_{\rm t}/\beta_{\rm d}$, where $\beta_{\rm d}$ depends on the stellar mass and age (\citealp{GRR2013}; see also \citealp{Law-Smith+2020,Ryu+2020ApJ}). Consequently, the true loss cone boundary is $J_{\rm lc,true}^2~(\simeq 2GMr_{\rm t}/\beta_{\rm d}\simeq J_{\rm lc}^2/\beta_{\rm d})$ rather than $J_{\rm lc}^2$. Therefore, full TDEs occurring in empty loss cone regime should satisfy $\beta\simeq J_{\rm lc}^2/J_{\rm lc,true}^2\simeq \beta_{\rm d}$. \citet{ZHL2023ApJ} demonstrated that in a realistic (but spherical) NSC, nearly all full TDEs in the $e<1$ quadrant should have $\beta\simeq 1$\footnote{Note that \cite{ZHL2023ApJ} used the simple definition of $r_{\rm t}$, so in their analysis $\beta_{\rm d}=1$.} (see the right panel of Figure 10 in that paper).

\begin{figure}
    \centering
    \includegraphics[width=0.27\linewidth]{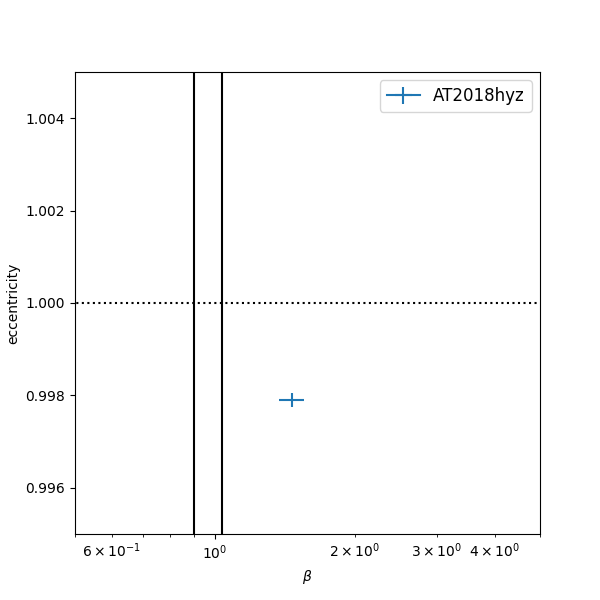}
    \includegraphics[width=0.27\linewidth]{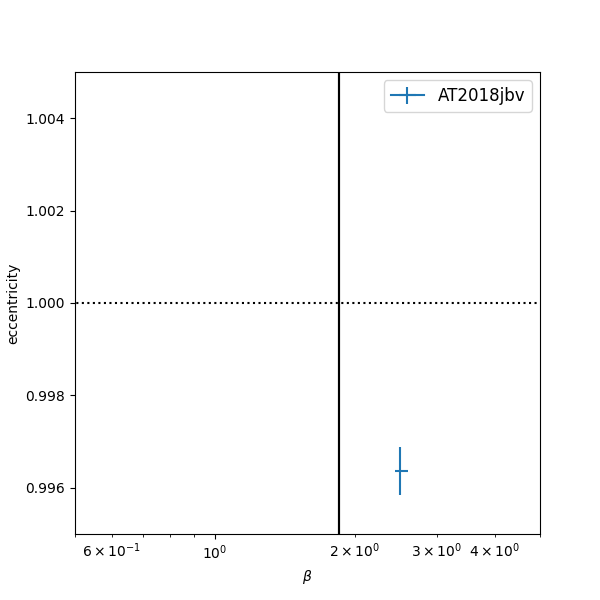}
    \includegraphics[width=0.27\linewidth]{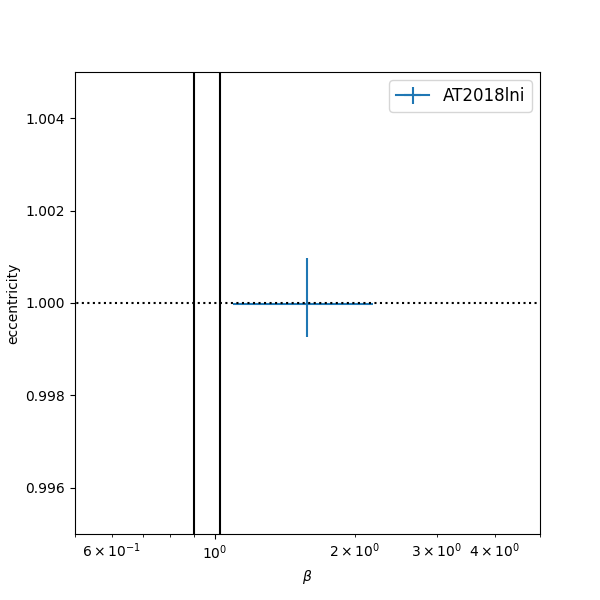}
    \includegraphics[width=0.27\linewidth]{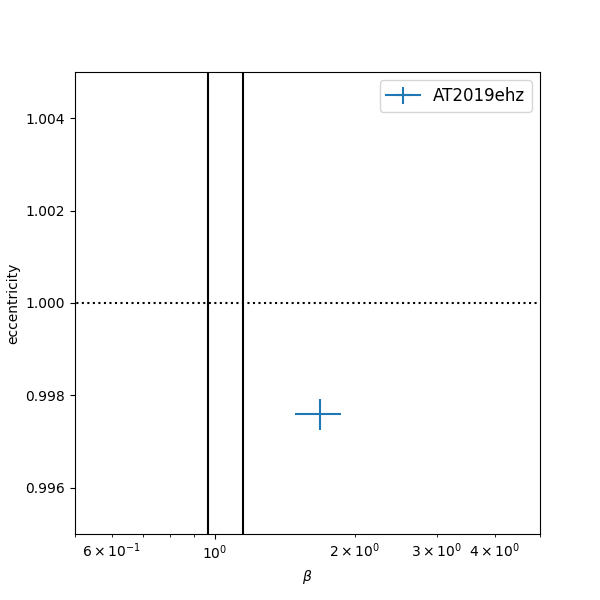}
    \includegraphics[width=0.27\linewidth]{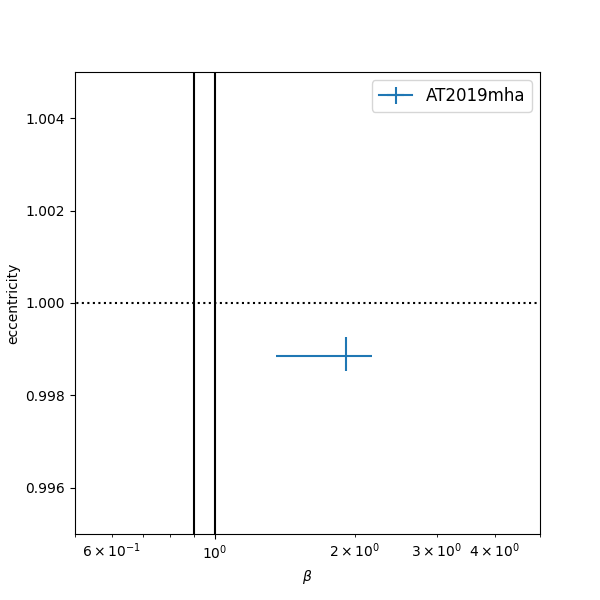}
    \includegraphics[width=0.27\linewidth]{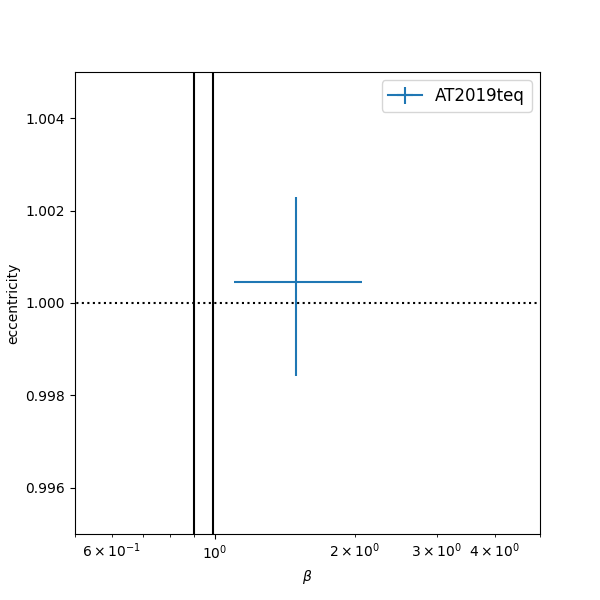}
    \includegraphics[width=0.27\linewidth]{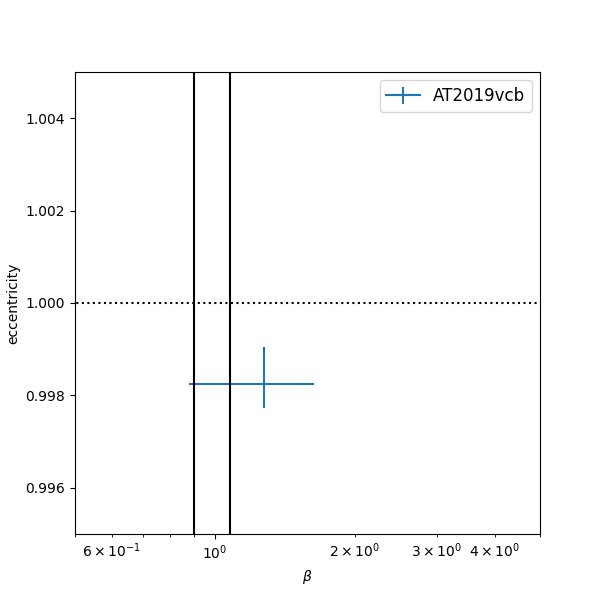}
    \includegraphics[width=0.27\linewidth]{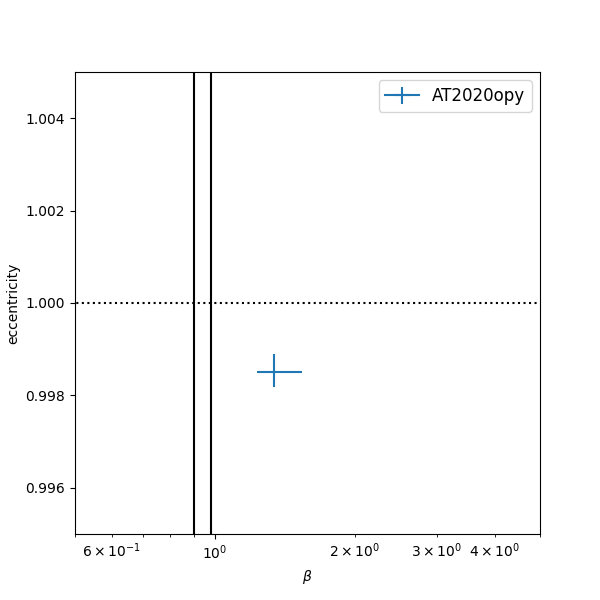}
    \caption{The positions of the 8 full TDEs with $\beta>\beta_{\rm d}$ in the $e$-$\beta$ plane. The vertical black lines indicate the range of $\beta_{\rm d}$, estimated from the posterior mass distribution of the disrupted star (see, equations 1--5 of \cite{Zhong2025}). The eccentricities of AT 2018lni and AT 2019teq are consistent with $e=1$ (albeit with large uncertainties), so they are permitted by the model of \cite{ZHL2023ApJ}. The other 6 TDEs lie beyond the boundary $\beta\simeq \beta_{\rm d}$ (for $e<1$) derived in \cite{ZHL2023ApJ}.
    }
    \label{fig:ecc-beta-FTDE}
\end{figure}

If TDEs are solely driven by two-body relaxation in spherical host NSCs, essentially no events should occupy the region with $e<1$ and $\beta>\beta_{\rm d}$. We identify 8 full TDEs with $\beta>\beta_{\rm d}$, and plot their positions in the $e$-$\beta$ plane in Figure~\ref{fig:ecc-beta-FTDE}. Since our model uses the fallback rate library from \cite{GRR2013}, $\beta_{\rm d}$ ranges from $0.9$ (for stars modeled as $\gamma=5/3$ polytrope) to $1.85$ (for $\gamma=4/3$ polytrope). The positions of AT 2018lni and AT 2019teq in the $e$-$\beta$ plane are consistent with production via two-body relaxation in spherical clusters. However, the remaining 6 TDEs must originate from other mechanisms.

In the work of \cite{CK1978}, loss cone refilling is driven by two-body relaxation, a cumulative effect of weak gravitational encounters among stars in a globular cluster. However, in NSC, the dense environment within the black hole's influence radius makes strong gravitational encounters non-negligible. Such encounters can induce large angular momentum changes, depositing stars deep within the loss cone and resulting in high-$\beta$ disruptions~\citep{WeissbeinSari2017}. Conversely, \citet{Teboul+2024} found that strong scattering can also eject stars from the cluster, thereby reducing the overall TDE rate.

The theoretical model of \citet{ZHL2023ApJ} assumes a spherically symmetric nuclear star cluster (NSC). However, real NSCs may exhibit deviations from spherical symmetry; for instance, bulk rotation has been observed in the Milky Way's NSC, leading to a flattened, axisymmetric structure~\citep{Feldmeier+2014AA}. In an axisymmetric potential, angular momentum is not conserved, allowing stars to experience larger variations in $\mathit{J}^2$ and penetrate deeper into the loss cone (i.e. achieve $\beta>\beta_{\rm d}$) compared to stars with the same orbital energy in a spherical cluster (see Fig. 8 of \citealt{VM2013}). Nonetheless, the condition $e<1$ implies that these six TDEs are bound to the black hole, meaning their maximum distances from the black hole remain within its influence radius. Within this region, the gravitational potential is dominated by the black hole and is effectively spherically symmetric. Therefore, axisymmetric NSCs may not fully account for these 6 events.

\section{Summary}
\label{SECT:Summary}
TDE is a promising method for detecting dormant MBHs and measuring their masses via light curve fitting. However, most existing studies rely on parabolic TDE models (e.g., using the widely adopted \texttt{MOSFiT} package). We argue that neglecting the orbital energy of the disrupted star in light curve modeling can introduce systematic biases in the derived black hole mass. To address this, we present a non-parabolic TDE model that modifies the debris mass distribution ($dm/d\epsilon$) based on the orbital energy, which is treated as a new fitting parameter $\tilde{\epsilon}_{\rm orb}$.

We fit the light curves of the 30 TDEs from \citetalias{HvVC2023} using both our non-parabolic TDE model (with $\tilde{\epsilon}_{\rm orb}$ as a free parameter) and the parabolic TDE model (fixing $\tilde{\epsilon}_{\rm orb}=0$). The comparison of the derived black hole masses reveals a trend consistent with the relation $M_{\rm BH} \propto (1-\tilde{\epsilon}_{\rm orb})^{17/5}$ derived in Section~\ref{SUBSEC:BH_mass_vs_Eorb}. Thus, if a TDE originates from a star on an eccentric (hyperbolic) orbit, fitting its light curve with a parabolic model systematically underestimates (overestimates) $M_{\rm BH}$.

We also measure, for the first time, the orbital eccentricity of these 30 TDEs. Previous studies based on parabolic models cannot provide eccentricity information due to the assumption $e=1$. With this data, we can explore the origin of the disrupted stars by mapping the events in the $e$-$\beta$ plane. We find that all but six TDEs are consistent with production via two-body relaxation in a spherical nuclear cluster, as they lie within the permitted region derived by \cite{ZHL2023ApJ}. The origin of the six outliers remains unclear and warrants further investigation.

\normalem
\begin{acknowledgements}
This work is supported by the National Key Research and Development Program of China (No. 2024YFA1611603) and the Yunnan Key Laboratory of Survey Science (No. 202449CE340002). We thank the participants of the TDE FORUM (Full-process Orbital to Radiative Unified Modeling) online seminar series for their inspiring discussions.

\end{acknowledgements}
  
\bibliographystyle{raa}
\bibliography{ZTF_TDE}

\end{document}